\documentclass[useAMS,usenatbib,usegraphicx]{temp}
\usepackage{amsmath}
\usepackage[mathlines]{lineno}

\newcommand{\bI}{\bmath{I}}
\newcommand{\bone}{\bmath{1}}
\newcommand{\bmu}{\bmath{\mu}}
\newcommand{\mcM}{\mathcal{M}}
\newcommand{\bQ}{\bmath{Q}}
\newcommand{\bR}{\bmath{R}}
\newcommand{\bs}{\mathbf{s}}
\newcommand{\bx}{\mathbf{x}}
\newcommand{\bep}{\boldsymbol{\varepsilon}}
\newcommand{\bSigma}{\boldsymbol\Sigma}

\newcommand{\N}{\text{N}}

\newcommand{\lb}{\left[}
\newcommand{\rb}{\right]}
\newcommand{\lbr}{\left\{}
\newcommand{\rbr}{\right\}}
\newcommand{\lp}{\left(}
\newcommand{\rp}{\right)}
\newcommand{\bmat}{\begin{pmatrix}}
\newcommand{\emat}{\end{pmatrix}}

\title[Clustering polar bears]{Accounting for phenology in the analysis of animal movement}

\author{Henry R. Scharf$^{1,*}$\email{henry.scharf@colostate.edu},
  Mevin B. Hooten$^{2, 1}$, 
  Ryan R. Wilson$^{3}$, 
  George M. Durner$^{4}$, and  
  Todd C. Atwood$^{4}$\\
  $^{1}$Department of Statistics,
  Colorado State University,
  Fort Collins, CO, USA \\
  $^{2}$U.S. Geological Survey,
  Colorado Cooperative Fish and Wildlife Research Unit, \\
  Department of Fish, Wildlife, and Conservation Biology,
  Department of Statistics, \\
  Colorado State University,
  Fort Collins, CO, USA \\
  $^{3}$U.S. Fish and Wildlife Service,
  1011 E Tudor Road,
  Anchorage, AK 99503, USA \\
  $^{4}$U.S. Geological Survey, 
  Alaska Science Center, 
  4210 University Drive,
  Anchorage, AK 99508, USA
}

\begin{document}
% \linenumbers
%  This will produce the submission and review information that appears
%  right after the reference section.  Of course, it will be unknown when
%  you submit your paper, so you can either leave this out or put in
%  sample dates (these will have no effect on the fate of your paper in the
%  review process!)

\date{{\it March} 2018. {\it December} 2018.  {\it
February} 2019.}

%  These options will count the number of pages and provide volume
%  and date information in the upper left hand corner of the top of the
%  first page as in published papers.  The \pagerange command will only
%  work if you place the command \label{firstpage} near the beginning
%  of the document and \label{lastpage} at the end of the document, as we
%  have done in this template.

%  Again, putting a volume number and date is for your own amusement and
%  has no bearing on what actually happens to your paper!

\pagerange{\pageref{firstpage}--\pageref{lastpage}}
\volume{00}
\pubyear{2019}
\artmonth{February}

%  The \doi command is where the DOI for your paper would be placed should it
%  be published.  Again, if you make one up and stick it here, it means
%  nothing!

\doi{10.1111/j.1541-0420.2005.00454.x}

%  This label and the label ``lastpage'' are used by the \pagerange
%  command above to give the page range for the article.  You may have
%  to process the document twice to get this to match up with what you
%  expect.  When using the referee option, this will not count the pages
%  with tables and figures.

\label{firstpage}

%  put the summary for your paper here

\begin{abstract} 
The analysis of animal tracking data provides important scientific understanding and discovery in ecology. Observations of animal trajectories using telemetry devices provide researchers with information about the way animals interact with their environment and each other. For many species, specific geographical features in the landscape can have a strong effect on behavior. Such features may correspond to a single point (e.g., dens or kill sites), or to higher-dimensional subspaces (e.g., rivers or lakes). Features may be relatively static in time (e.g., coastlines or home-range centers), or may be dynamic (e.g., sea ice extent or areas of high-quality forage for herbivores). We introduce a novel model for animal movement that incorporates active selection for dynamic features in a landscape. Our approach is motivated by the study of polar bear (\textit{Ursus maritimus}) movement. During the sea ice melt season, polar bears spend much of their time on sea ice above shallow, biologically productive water where they hunt seals. The changing distribution and characteristics of sea ice throughout the year means that the location of valuable habitat is constantly shifting. We develop a model for the movement of polar bears that accounts for the effect of this important landscape feature. We introduce a two-stage procedure for approximate Bayesian inference that allows us to analyze over 300,000 observed locations of 186 polar bears from 2012--2016. We use our model to estimate a spatial boundary of interest to wildlife managers that separates two sub-populations of polar bears from the Beaufort and Chukchi seas.
\end{abstract}

%  Please place your key words in alphabetical order, separated
%  by semicolons, with the first letter of the first word capitalized,
%  and a period at the end of the list.
%

\begin{keywords}
animal movement; resource selection function; sea ice; \textit{Ursus maritimus}; 
\end{keywords}

%  As usual, the \maketitle command creates the title and author/affiliations
%  display

\maketitle

%  If you are using the referee option, a new page, numbered page 1, will
%  start after the summary and keywords.  The page numbers thus count the
%  number of pages of your manuscript in the preferred submission style.
%  Remember, ``Normally, regular papers exceeding 25 pages and Reader Reaction
%  papers exceeding 12 pages in (the preferred style) will be returned to
%  the authors without review. The page limit includes acknowledgements,
%  references, and appendices, but not tables and figures. The page count does
%  not include the title page and abstract. A maximum of six (6) tables or
%  figures combined is often required.''

%  You may now place the substance of your manuscript here.  Please use
%  the \section, \subsection, etc commands as described in the user guide.
%  Please use \label and \ref commands to cross-reference sections, equations,
%  tables, figures, etc.
%
%  Please DO NOT attempt to reformat the style of equation numbering!
%  For that matter, please do not attempt to redefine anything!

\section{Introduction}\label{sec:intro}
For decades, study of animal movement has led to important scientific understanding and discoveries in ecology \citep{Hootenbook2017}. Observations of animal trajectories using telemetry devices, such as radio collars, have provided researchers with information about the way animals interact with their environment (e.g., \citealt{Manly2002}; \citealt{Johnson2008}) and each other \citep[e.g.,][]{Niu2016, Scharf2016, Scharf2018}. In the case of the former, one of the most common approaches used by ecologists is to analyze the rates at which animals use certain types of habitats relative to the distribution of habitat types available to them (\citealt{Manly2002}; \citealt{Lele2006}; \citealt{Hootenbook2017}). Such analyses typically model the probability of an individual using a particular location as a weighted combination of all available locations \citep[e.g.,][]{Johnson2006, Northrup2013}. The weights are then referred to as the resource selection function (RSF) and provide insight into which portions of a landscape are most valuable to the study species. What constitutes an available location depends on the characteristics of the particular species under study and the rate at which telemetry observations are gathered. 

For many species, specific geographical features in the landscape can have a strong effect on where individuals choose to move \citep{Tracey2005}. Such features are sometimes well-summarized by a single point (e.g., dens or kill sites), but may also correspond to higher-dimensional subspaces (e.g., rivers or lakes). Their locations may be relatively static in time (e.g., coastlines or home range centers in \citealt{Brost2016}), or may be dynamic, potentially varying periodically (e.g., sea ice extent or areas of high-quality forage for herbivores). The term pheonology is used to refer to those characteristics of an ecosystem that demonstrate seasonal patterns of variation. While not always framed in the context of resource selection, these landscape features can nevertheless be thought of as resources, and the behavior of animals may demonstrate selection for (or against) points near the feature. We introduce a novel suite of models for the analysis of animal movement that incorporates active selection for features in a landscape that may have complex and dynamic shapes. 

Our modeling framework is motivated by the study of polar bear (\textit{Ursus maritimus}) movement. Polar bears spend much of their time on sea ice over shallow, biologically productive water where they hunt ringed seals (\textit{Pusa hispida}). The distribution of this primary food source can be highly variable and is known to depend on a wide variety of factors including season, bathymetry, and various characteristics of the sea ice \citep[e.g.,][]{Kelly2010}. Ringed seals depend heavily on sea ice throughout their lives and use both fast ice (ice that is connected to land) and pack ice (ice that is free to drift) \citep[e.g.,][]{Finley1983, Wiig1999} for feeding, rearing pups, and molting. A consistent pattern across multiple studies is that some of the most valuable habitat for ringed seals, and hence polar bears, is near the interface between sea ice and the ocean \citep{Frost2004, Durner2009, Crawford2012, Rode2015, Atwood2016} especially during the sea ice melt season, the part of the year when sea ice first breaks up and contracts toward the pole, then freezes and expands southward again. The changing distribution and characteristics of sea ice throughout the late spring through early fall means that the location of valuable ice-edge habitat is constantly shifting.

As climate change alters the rate at which sea ice thaws and freezes, as well as the size of its minimum and maximum extents, there is increasing concern about how polar bears are responding to these dramatic shifts in their environment \citep{Rode2014}. Our goal is to develop a model for the movement of polar bears that explicitly accounts for the effect of the changing sea ice and can be incorporated into a wide variety of hierarchical models used to better understand polar bear ecology. In Section~\ref{sec:application}, we use our model to estimate a spatial boundary of interest to wildlife managers that separates two sub-populations of polar bears from the Beaufort and Chukchi seas.

\section{Model Development}\label{sec:model}
\subsection{Feature preference}
To account for an individual's preference for areas in a landscape near (or far from) a particular feature of interest, we take a use-availability approach. We define the selection weight of a particular location to be a parametric function of the distance from the location to a feature of interest on the landscape. Estimates of the relevant parameters provide a summary of how strongly the feature affects the behavior of observed individuals. We model availability similar to \cite{Hjermann2000}, \cite{Christ2008}, \cite{Johnson2008a}, and \cite{Brost2015} who used radial distributions centered on the most recently-observed location to define the continuously-valued availability at each point in time.

Let $\bmu(t)$ be the location of an individual at time $t \in \lbr 1, \dots, T \rbr$, where we assume for now that observation times are equally spaced with no missing values. We define the conditional probability density for $\bmu(t)$ as
\begin{linenomath*} 
\begin{equation}
  \begin{array}{ll}
    \lb \bmu(t) | \bmu(t-1), \sigma_\mu^2, \mcM(t), \tau^2 \rb \propto \\
      \quad \N\!\lp \bmu(t); \bmu(t-1), \sigma^2_\mu \bI_2 \rp^{\bone_{t > 1}} g(\bmu(t); \mcM(t), \tau^2)
  \end{array}\label{eqn:movement}
\end{equation}
\end{linenomath*} 
where we use square brackets to denote a probability density, and $\N\!\lp \bx; \bmu, \bSigma \rp$ denotes the normal probability density function with mean $\bmu$ and variance $\bSigma$ evaluated at $\bx$. For $t>1$, the conditional distribution is proportional to the product of two components, the first of which is the density of a bivariate Gaussian distribution centered on the previous location of the individual, and defines the availability of each point on the landscape as in \cite{Christ2008}. The availability component induces positive auto-correlation in the joint process $\bmu \equiv \lp \bmu(1), \dots, \bmu(T) \rp'$, with larger values of $\sigma_\mu^2$ resulting in processes with greater distances between consecutive locations and faster, more erratic movement. 

The function $g$ is a RSF and controls the effect a particular feature in the landscape has on an individual's movement. Let $\mcM(t)$ denote the set of points that comprise the feature of interest (e.g., the interface between sea ice and ocean). We define the function $g$ as
\begin{linenomath*} 
\begin{equation}
  g(\bmu(t); \mcM(t), \tau^2) = \exp \lbr
  -\min_{\bx \in \mcM(t)} \| \bmu(t) - \bx \|^2_2 / 2\tau^{2} \rbr,
  \label{eqn:g}
\end{equation}
\end{linenomath*} 
(where $\| \cdot \|_2$ is the $\ell_2$ or Euclidean norm) so that the value of $g$ is highest near $\mcM(t)$, and drops to zero as $\bmu(t)$ moves away from $\mcM(t)$. The value of $\tau^2$ controls the range at which $g$ effectively reduces to zero. We show in Section~\ref{sec:linearization} why this particular parametric form for the RSF can be used to leverage computational efficiencies in parameter estimation. 

In practice, it will often be useful to define $g$ such that it achieves its largest values at locations $\bmu(t)$ near $\mcM(t)$ so that the conditional density given in~(\ref{eqn:movement}) has probability mass concentrated near $\mcM(t)$. Specifying $g$ in this way provides a method for modeling movement that exhibits preference for the region of the landscape near the feature of interest. 

The model for the discrete-time continuous-space process $\bmu$ provides a useful tool for modeling the movement of an individual responding to a one-dimensional feature on a landscape. In Section~\ref{sec:application}, we apply the model to the movement of polar bears with the ultimate goal of clustering individuals into disjoint sub-populations based on space use. By including availability and resource selection as part of a larger hierarchical structure, we are able to account for polar bears preference for habitats that facilitate the depredation of seals, which, if ignored, might result in biased inference about sub-population membership.

The conditional density in~(\ref{eqn:movement}) is only defined up to a constant of proportionality, \linebreak$\int \N\!\lp \bmu(t); \bmu(t-1), \sigma^2_\mu \bI_2 \rp^{\bone_{t > 1}} g(\bmu(t); \mcM(t), \tau^2) d\bmu(t)$, that must be computed as part of any likelihood-based estimation procedure. We employ a Bayesian hierarchical methodology and fit our model for polar bear movement using Markov chain Monte Carlo (MCMC), which requires computation of the normalizing constant several times at each iteration of the algorithm. For a general feature, $\mcM(t)$, the normalization constant is not analytically tractable. Thus, some form of numerical integration is required to fit the model to data, the computational cost of which precludes such an approach in our application. In the RSF literature, the normalization constant has typically been approximated using either a coarse spatial discretization (e.g., \citealt{Warton2010}; \citealt{Brost2015}), or a randomized scheme based on an ``availability sample'' \citep[e.g.,][]{Northrup2013}. In the next section, we introduce a novel and computationally efficient approach for computing the necessary normalization constant that leverages conjugacy in the distributional form for $\bmu(t)$.

\subsection{Linearization approximation}\label{sec:linearization}
We implement a novel approximation technique that assumes locally linear structure in the shape of $\mcM(t)$, allowing for efficient approximation of the true conditional density of $\bmu(t)$. To motivate our approximation, we note that, for the special case when $\mcM(t)$ is a straight line, the RSF as defined in~(\ref{eqn:g}) can be written in a form similar to that of a bivariate Gaussian density function with a rank-deficient covariance matrix. 

\subsubsection{RSF for straight lines}\label{sec:RSF_straight_lines}
First, we consider the case of a vertical line, $\widetilde{\mcM}(t)$, in the real plane so that $\widetilde{\mcM}(t) \equiv \lbr (x, y) \in \mathcal{R}^2:x=h \rbr$. For this case, we have
\begin{linenomath*} 
\begin{equation*}
  \begin{array}{ll}
    g(\bmu(t); \widetilde{\mcM}(t), \tau^2) = \\ \quad 
      \exp \lbr -\frac{1}{2} \lp \bmu(t) - (h, y)' \rp' 
        \bQ(\tau^2) \lp \bmu(t) - (h, y)' \rp \rbr, \\
    \bQ(\tau^2) \equiv \begin{pmatrix} \tau^{-2} & 0 \\ 0 & 0 \end{pmatrix}
  \end{array}
\end{equation*}
\end{linenomath*} 
for all real-valued $y$. To allow for $\widetilde{\mcM}(t)$ that are straight, but not necessarily vertically oriented, we rotate the coordinate system through an angle $\theta$. Let $\bR(\theta)$ be the rotation matrix defined as $ \bR(\theta) \equiv \begin{pmatrix} \cos \theta & -\sin \theta \\
\sin \theta & \cos \theta \end{pmatrix},$ and let $\theta$ be defined such that $\bR'(\theta) \widetilde{\mcM}(t) = \lbr (x, y) \in \mathcal{R}^2:x=h \rbr$ for some real-valued $h$. Note that the inverse of a rotation matrix, $\bR^{-1}(\theta) = \bR(-\theta)$, is also equal to its transpose, $\bR'(\theta)$. The RSF defined in~(\ref{eqn:g}) is invariant under rigid transformations such as rotations, therefore 
\begin{linenomath*} 
\begin{equation*}
  \begin{array}{ll}
    g(\bmu(t); \widetilde{\mcM}(t), \tau^2) 
      = g(\bR'(\theta) \bmu(t); \bR'(\theta) \widetilde{\mcM}(t), \tau^2) \\
    = \exp \lbr -\frac{1}{2} \lp \bmu(t) - (h, y)' \rp' \bR(\theta) \bQ(\tau^2) 
      \bR'(\theta) \lp \bmu(t) - (h, y)' \rp \rbr.
  \end{array}
\end{equation*}
\end{linenomath*} 

\subsubsection{Linearizing complex landscape features}
The resulting form for $\lb \bmu(t)|\bmu(t-1), \widetilde{\mcM}(t) \rb$ is proportional to the product of two Gaussian distributions, one of which is improper. Provided $0 < \tau^2$, the product is a proper bivariate Gaussian distribution with mean, $\bmu^* \equiv \bmath\Sigma^* \lp \sigma_\mu^{-2} \bmu(t-1) + \bR(\theta) \bQ(\tau^2) \bR'(\theta) \widetilde{\bmath{m}} \rp$ and covariance, $\bmath\Sigma^* \equiv \lp \sigma_\mu^{-2} \bI_2 + \bR(\theta) \bQ(\tau^2) \bR'(\theta) \rp^{-1}$, where $\widetilde{\bmath{m}}$ is any point in $\widetilde{\mcM}(t)$. The distributional form of $\bmu(t)|\bmu(t-1), \widetilde{\mcM}(t)$ implicitly defines the appropriate normalization constant in~(\ref{eqn:movement}). Thus, if there exists some straight line $\widetilde{\mcM}(t)$ that represents a close approximation to $\mcM(t)$ near $\bmu(t-1)$, then $g(\bmu(t); \widetilde{\mcM}(t), \tau^2)$ may provide a reasonable approximation for $g(\bmu(t); \mcM(t), \tau^2)$ that alleviates the computational burden of repeatedly calculating the necessary normalization constant. A natural candidate for $\widetilde{\mcM}(t)$ is the line that is tangent to $\mcM(t)$ at the point on $\mcM(t)$ closest to $\bmu(t-1)$, because this is the portion of the feature most relevant to the conditional distribution of $\bmu(t)$.
\begin{figure}
  \centering
  \includegraphics[width=0.5\textwidth]{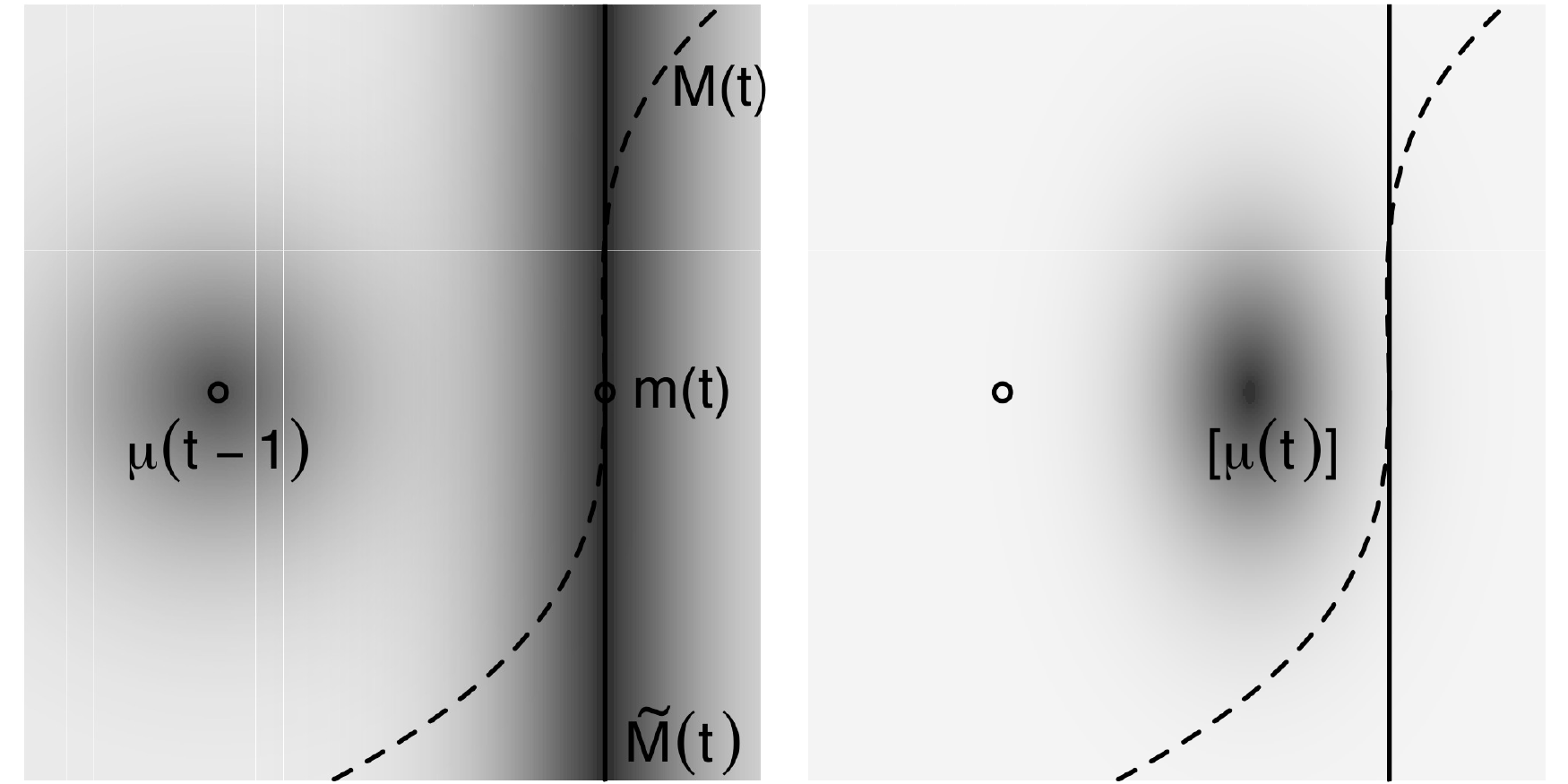}
  \caption{The product of the two densities in the left plot (one of which is improper) results in the density shown in the right plot. In the context of polar bear movement, the two densities in the left plot represent the location of an individual on day $t-1$ (point) and the RSF defined by the linearized boundary between, for example, the sea ice and the ocean (diagonal line). The true boundary is represented by the dashed curve. The density shown in the right plot represents the conditional density for the true location on day $t$.}
  \label{fig:linear_example}
\end{figure}
Let $\bmath{m}(t)$ be the point in $\mcM(t)$ nearest to $\bmu(t-1)$, and let $\widetilde{\mcM}(t)$ be the set of points that lie on the line tangent to $\mcM(t)$ at $\bmath{m}(t)$. Figure~\ref{fig:linear_example} shows a schematic illustrating the way the product of the two Gaussian components in~\eqref{eqn:movement} combine to result in a third Gaussian distribution. For $\widetilde{\mcM}(t)$ to result in an adequate approximation of the RSF, the linearized feature need only resemble the true feature in the vicinity of $\bmu(t)$. Outside of the immediate neighborhood, the availability distribution will be essentially zero, reducing the impact of errors in the RSF approximation. We discuss the appropriateness of the linearization approximation in greater detail and offer practical guidance for its use in future studies in Supporting Information~D.

\section{Application}\label{sec:application}
\subsection{Goals and previous work}
\begin{figure}
  \centering
  \includegraphics[width = 0.5\textwidth]{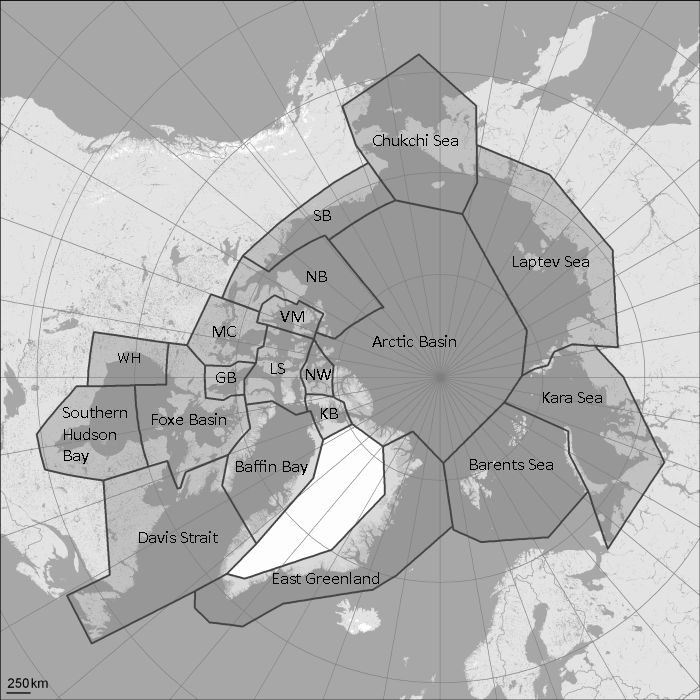}
  \caption{Sub-population boundaries of polar bears \citep{Obbard2010, U.S.FishandWildlifeService2016}. Sub-population abbreviations are: Southern Beaufort Sea (SB), Northern Beaufort (NB), Kane Basin (KB), Norwegian Bay (NW), Lancaster Sound (LS), Gulf of Boothia (GB), M'Clintock Channel (MC), Viscount Melville Sound (VM), and Western Hudson Bay (WH).}
  \label{fig:all_populations}
\end{figure}

There are a total of 19 recognized polar bear sub-populations in the circumpolar Arctic \citep[Figure~\ref{fig:all_populations},][]{Obbard2010}. However, the boundaries that delineate the sub-populations are challenging to precisely define because there are few barriers to movement for polar bears, and the changing extent and drift of the sea ice leads to periods of the year when individuals from different sub-populations may use overlapping portions of the landscape. Nevertheless, there are important reasons to determine a clear delineation of the sub-population boundaries. For example, wildlife management agencies such as the U.S. Fish and Wildlife Service (USFWS) use sub-population boundaries to help guide management decisions for polar bears, which are currently listed as `threatened' under the Endangered Species Act \citep{U.S.FishandWildlifeService2016}. There is also evidence that polar bears from different sub-populations are responding to climate change with differing degrees of success \citep{Rode2014, Ware2017}. Statistical methods that make use of polar bear movement data have been used to estimate these boundaries in the past \citep{Amstrup2005}. In what follows, we focus on demonstrating how our model may be used for estimating a sub-population boundary between the Chukchi Sea (CS) and Southern Beaufort Sea (SB) sub-populations. We compare previous statistical approaches with our proposed methodology in Supporting Information~A.

\subsection{Movement process model}
With the exception of mothers denning on land, most polar bears remain on the ice in the summer, following it north as it retreats away from continental land masses. Polar bears are specialist carnivores that use areas near the interface of sea ice and ocean to hunt seals during the sea ice melt season \citep{Durner2009}. However, some polar bears do remain on land through the warmest part of the year and, in fact, \cite{Atwood2016} concluded that there is evidence the number of individuals exhibiting this behavior is increasing. 

Two of the primary characteristics of polar bear movement are a tendency for most individuals to prefer portions of the landscape near the edge of the sea ice \citep{Durner2004, Ware2017}, and a tendency for individuals to occupy a general spatial region corresponding to the particular unobserved sub-population to which an individual polar bear is a member. To address the effect of the sea ice, we specify a model for movement that incorporates preference for areas near the sea ice boundary as it changes in time. We account for polar bears that remain on land during the summer by also including coastline as a feature on the landscape associated with increased rates of use, as individuals that spend the summer on land tend to remain on islands and/or near the coastline for most of the season \citep{Rode2015}. By accounting for habitat variability, we hope to avoid inadvertently clustering bears into sub-populations that confound true sub-population spatial regions with movement responding to changing sea ice. 

We take a Bayesian hierarchical modeling perspective, allowing us to specify models for both measurement error and the true unobserved movement process in a single coherent framework. We use a discrete-time approach to specify a model for polar bear movement, in which the conditional probability density for the location of an individual is proportional to a product of components corresponding to resource availability and selection. The discrete-time approach is motivated by the scale at which we have measurements of Arctic sea ice, which features as a direct effect in our model for movement. We use estimates of the extent of sea ice provided by the National Snow and Ice Data Center \citep{Fetterer2010} that are available at daily intervals at 4km resolution; thus, we model movement as a discrete-time stochastic process on a daily scale. 

Let $\bmu_i(t)$ denote the location of individual $i \in \lbr 1, \dots, N \rbr$ at time $t \in \mathcal{T}_i$, where $\mathcal{T}_i$ is a consecutive set of times from the reference set $\mathcal{T} = \lbr 1, \dots, T \rbr$, and let $z_i$ be a binary random variable that equals 1 if individual $i$ belongs to the CS sub-population, and 0 if it belongs to the SB sub-population (for the general case of more than two sub-populations, $z_i$ may be specified as a categorical variable coming from a multinomial distribution). Additionally, we denote by $\mcM(t)$ the set of points in the plane defined by the union of coast line, and the edge of the sea ice. Figure~5 in Supporting Information~A shows realizations of the dynamic feature of interest, $\mcM(t)$, for three days in 2018. We model the conditional distribution of $\bmu_i(t)$ as
\begin{linenomath*} 
\begin{equation}
  \begin{array}{ll}
    \lb \bmu_i(t) | \bmu_i(t - 1), \sigma_\mu^2, \bmu_{\text{CS}}, \bmu_{\text{SB}},
    \bSigma_{\text{CS}}, \bSigma_{\text{SB}}, z_i, \tau^2 \rb \propto \\ \quad
    \underbrace{\N \! \lp \bmu_i(t); \bmu_{\text{CS}}, \bSigma_{\text{CS}} \rp^{z_i}
    \N \! \lp \bmu_i(t); \bmu_{\text{SB}}, \bSigma_{\text{SB}} \rp^{1-z_i}}_{\text{(i)}}
    \\ \qquad \times \;
    \underbrace{\N \! \lp \bmu_i(t); \bmu_i(t-1), \sigma^2_\mu \bI_2 \rp^{\bone_{t > \min(\mathcal{T}_i)}}}_{\text{(ii)}} 
    \\ \qquad \times \;
    \underbrace{g(\bmu_i(t); \mcM(t), \tau^2)^{\bone_{t > \min(\mathcal{T}_i)}}}_{\text{(iii)}}
    \label{eqn:application}
  \end{array}
\end{equation}
\end{linenomath*} 

Each component in~(\ref{eqn:application}) captures a different feature of the movement process. Namely, these are (i) the association of each individual bear with a sub-population-level central place, (ii) the temporal dependence between locations on consecutive days, and (iii) a RSF that appropriately weights locations near a coastline or the edge of the sea ice. The first two terms can be thought of as a two-component availability function that incorporates a sub-population activity center and movement constraints, similar in many respects to the modeling specification of \cite{Christ2008} and \cite{Johnson2008a}. The third term is a RSF that models the preference polar bears exhibit for habitat near either a coastline, or the sea ice boundary. The exponents in components (ii) and (iii) are indicator functions that remove those effects at the initial time point for which we assume the only information we have about an individual's location comes from their association with a particular sub-population.

\subsection{Activity centers and sub-population membership}

The second component in~(\ref{eqn:application}), for $t > \min(\mathcal{T}_i)$, and first component for $t= \min(\mathcal{T}_i)$ is a bivariate Gaussian distribution centered on one of two central places, $\bmu_{\text{SB}}$ and $\bmu_{\text{CS}}$, corresponding to the centers of the SB and CS sub-populations, respectively. The $2\times2$ covariance matrices, $\bSigma_{\text{CS}}$ and $\bSigma_{\text{SB}}$, control the strength of the effect the sub-population center has on the movement of each individual. As the marginal variances of $\bSigma_{\text{CS}}$ and $\bSigma_{\text{SB}}$ increase, individuals are allowed to range farther from their central place. We specify prior distributions for $\bSigma_{\text{CS}}$ and $\bSigma_{\text{SB}}$ based on their spectral decompositions. Details are provided in Supporting Information~B.

The binary random variables, $z_i$, indicate the particular sub-population with which individual $i$ associates. We specify Bernoulli prior distributions for each $z_i$ with probability 0.5, corresponding to balanced \textit{a priori} classification of each individual.

\subsection{Time-varying RSF}\label{sec:time_varying_RSF}
For the application to polar bear movement, we modified the RSF introduced in~\eqref{eqn:g} slightly to account for an important aspect of polar bear ecology. Polar bears' preference for habitat near the coastlines and boundary between ocean and sea ice can vary seasonally throughout the year. Previous analyses of polar bear movement concluded that the preference for habitat near the sea ice boundary is highest during summer and autumn months and weakest in the spring \citep{Durner2004, Durner2009, Wilson2016}. Thus, we defined a RSF with the same shape characteristics as in~\eqref{eqn:g}, but with an added indicator function that accounts for the difference in seasons when the landscape feature is most relevant, and when it has negligible impact on movement. Let $a$ and $b$ denote the days of the year $(0,\dots,364)$ on which this ``summer'' season begins and ends, respectively. Then the modified RSF was defined as
\begin{linenomath*} 
\begin{equation}
  g(\bmu_i(t); \mcM(t), \tau^2) = \bone_{a < t < b}
    \exp \lbr -\min_{\bx \in \mcM(t)} \| \bmu(t) - \bx \|^2_2 / 2\tau^{2} \rbr. \label{eqn:g_mod}
\end{equation}
\end{linenomath*} 
We specified Gaussian prior distributions for the end-points of the ``summer'' such that $a \sim \N(135, 14^2)$, $b \sim \N(319, 14^2)$, corresponding to prior start and end date distributions with means of May 15 and November 15, respectively, and a standard deviation of two weeks. 

One way to interpret this generalization of~\eqref{eqn:g} is to define the RSF in the original way, but let $\tau^2$ vary in time. The modified RSF given by~\eqref{eqn:g_mod} arises for the case of a dynamic $\tau^2$ parameter that is constant and finite during the ``summer'' season, and infinite during the rest of the year. Thus, outside of the season defined by $(a, b)$ the landscape feature defined by $\mcM(t)$ has no effect on movement.

\subsection{Measurement error}\label{sec:measurement_error}
We analyzed telemetry observations of $N=186$ polar bears made by the USFWS (USFWS unpublished data) and the U.S. Geological Survey (USGS) (USGS unpublished data) using a variety of different telemetry device types contaminated with measurement error of varying severity. All locations were projected to a grid using an Albers equal area projection, with all distances measured in kilometers. In each case, we model the observed locations as centered on the true, unobserved location of the individual with an additive measurement error process as $\bs_i(t^*) = \bmu_i(t^*) + \bep_i(t^*), \; t^* \in \lp 1, T \rp$, where the distribution for $\bep_i(t^*)$ depends on the particular device used to make the observation $\bs_i(t^*)$, and we allow measurements to be made at any point on the continuous interval $\lp 1, T \rp$. We provide full details related to the measurement error model and address the misalignment between this continuous-time scale and the discrete scale used to model the movement process in Supporting Information~B.

\subsection{Results}\label{sec:results}
\subsubsection{Sub-populations}
\begin{figure}
  \centering
  \includegraphics[width = 0.5\textwidth]{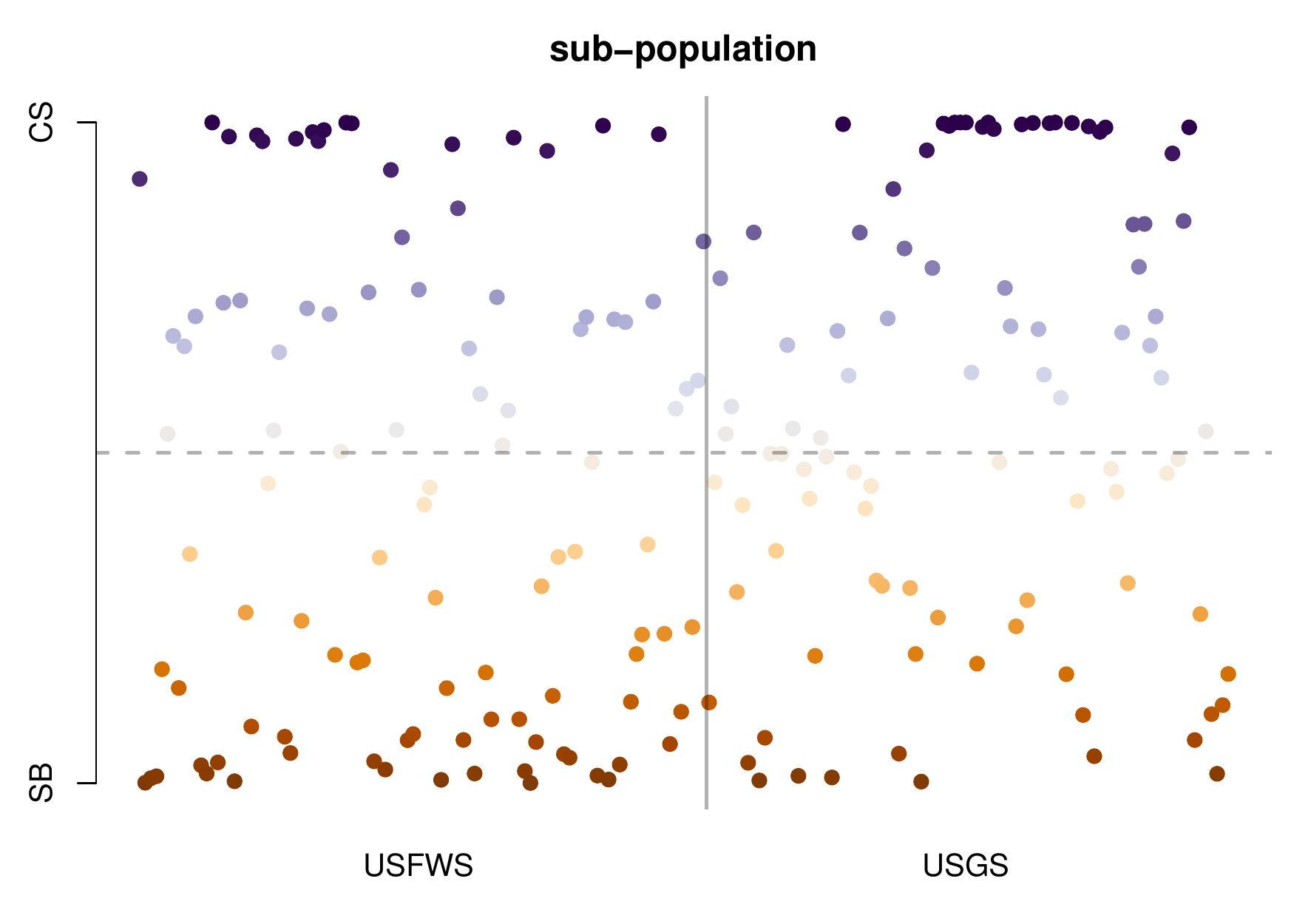}
  \caption{Posterior means for each $z_i$, organized by the agency responsible for tagging the individual. Darker points reflect posterior means closer to 0 (Southern Beaufort Sea; SB) and 1 (Chukchi Sea; CS). This figure appears in color in the electronic version of this article.}
  \label{fig:sub_pop}
\end{figure}

Using the two-stage approach, we fit the model to all observations made by USFWS and USGS from 2012--2016 (186 unique individuals). Table~\ref{tab:CI} gives the posterior medians and equal-tailed 95\% credible intervals for each variance parameter, as well as relevant prior distributions and hyperpriors. Figure~\ref{fig:sub_pop} shows the posterior mean of each $z_i$ grouped by the agency that tagged the individuals. Posterior means of the class indicator variables can be interpreted as the posterior probability that individual $i$ is a member of the Chukchi Sea sub-population. 
\begin{table}
  \caption{Posterior medians and equal-tailed credible intervals for all model parameters, as well as prior distributions and hyper-parameters.}
  \label{tab:CI}
  \begin{center}
    \begin{tabular}{r|rc|c}
      \Hline 
      \multicolumn{1}{c}{} & \multicolumn{2}{c}{posterior summary} & prior \\
      \hline
      parameter       & median & (2.5\%, 97.5\%)     & density \\
      \hline                                                                         
      $\sigma_\mu^2$  & 272   & (267, 277)        & $\text{IG}(6, 1125)$  \\
      $\tau^2$        & 8600 & (8000, 9500)    & $\text{IG}(6, 32000)$  \\
      $\sigma_\mu$  & 16.5   & (16.3, 16.6)        &   \\
      $\tau$        & 93 & (89, 97)    &   \\
      $z_i$ & \multicolumn{2}{c}{see Figure~\ref{fig:sub_pop}} & $\text{Bern}(0.5)$ \\
      $a$             & 69    & (28, 83)          & $\N(135, 14^2)$\\
      $b$             & 337    & (327, 349)          & $\N(319, 14^2)$ \\
      \hline
    \end{tabular}
  \end{center}
\end{table}

To produce a meaningful spatial delineation of the two sub-populations from which our study animals were drawn, we use a derived quantity related to the inferred locations of the sub-population activity centers. If we consider the observation of a single new location, $\bmu_{J+1}(t)$, and integrate across all arbitrary features, effectively removing the effect of the RSF, it can be shown that the posterior probability that $z_{J+1} = 1$ is given by
\begin{linenomath*} 
\begin{equation*}
\begin{array}{ll}
  \text{Pr}\lp z_{J+1} = 1 | \bmu_{J+1}(t), \bmu_{\text{CS}}, \bmu_{\text{SB}}, \bSigma_{\text{CS}}, \bSigma_{\text{SB}} \rp \propto 
  \\ \quad
  \frac{\N(\bmu_{J+1}(t); \bmu_{\text{CS}}, \bSigma_{\text{CS}})}{\N(\bmu_{J+1}(t); \bmu_{\text{SB}}, \bSigma_{\text{SB}})}.
\end{array}
\end{equation*}
\end{linenomath*} 
For the case of an uninformative prior, $p_{J+1} = 0.5$, the contour corresponding to $\text{Pr}(z_{J+1} = 1) = 0.5$ is defined by the points in the plane where the two normal densities are equal. It can be shown that these contours are the roots of quadratic polynomials in two dimensions. By computing the contour at each iteration in the MCMC algorithm, we can obtain draws from the posterior distribution, a summary of which provides wildlife managers with a way to delineate the boundary between the two sub-populations.

\begin{figure*}
\includegraphics[width=\textwidth]{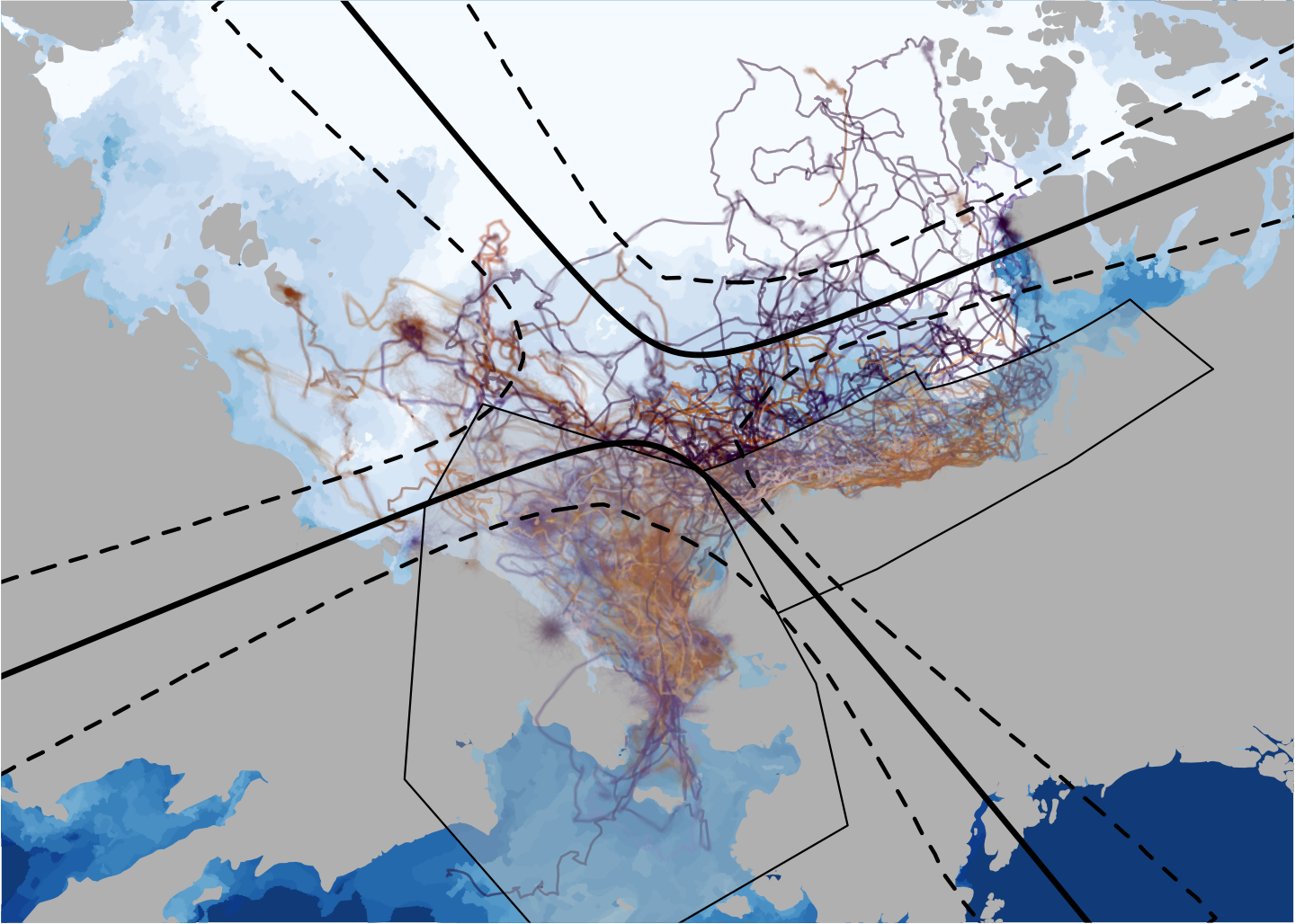}
\caption{The orange and purple lines represent trajectories drawn from the process imputation distributions (Supporting Information~B) of all polar bears from 2012--2016. The colors correspond to the posterior means of $z_i$, with dark orange hues corresponding to values close to 0 (Southern Beaufort Sea), white corresponding to values close to 0.5, and dark purple corresponding to values close to 1 (Chukchi Sea). Weekly measurements of sea ice extent for March-September 2016 are shown as light blue polygons, with the darkest polygon corresponding to open ocean, the second darkest polygon to March 1st, and the lightest polygon to September 30th. The black line shows central-measure summary of the distribution for the derived spatial boundary where, marginally, the probability of sub-population membership is balanced (Section~\ref{sec:results}). The dashed lines show pointwise equal-tailed 95\% credible intervals orthogonal to the central-measure boundary. The large polygons with thin black borders show the current sub-population delineations (CS and SB from left to right; see Figure~\ref{fig:all_populations}).}
\label{fig:map}
\end{figure*}

Figure~\ref{fig:map} shows a map of the region encompassing the Chukchi and southern Beaufort seas. The solid black line corresponds to a central-measure summary of the posterior distribution of the derived spatial boundary, and the dashed lines show equal-tailed 95\% pointwise credible intervals computed orthogonal to the solid line (more details are given in Supporting Information~C). The portion of the inferred boundary most relevant in this application is in the bottom right quarter of the map and suggests at most a small shift from the currently accepted sub-population delineation, denoted by the large polygons with thin black lines (also shown in Figure~\ref{fig:all_populations}). The discrepancy between the currently used sub-population boundary and the one we derive in this paper is likely due to a combination of factors including both the use of novel statistical methodology, and systematic changes to polar bear behavior as sub-populations respond to rapidly changing habitat conditions. We also note that a potentially informative landscape feature, ocean depth, has been included in past RSF analyses of polar bears but was not accounted for in this analysis. A goal of future statistical and ecological research is to develop methods that can account for both the sea ice boundary and landscape features like ocean depth.

\subsubsection{Response to habitat characteristics}
The posterior distribution of the parameters defining the ``summer'' season suggests that the feature $\mcM(t)$, defined as the union of the interface between sea-ice and ocean and the continental coastlines, has the greatest impact on the behavior of polar bears between March 10 and December 3. While the end of the season aligns closely with prior expectations, there appears to be evidence in the data that polar bears show a preference for habitat near either coastlines or the edge of the sea ice more than two months earlier than expected. However, we note that because of the way telemetry devices were deployed, the posterior distribution for $a$ must be interpreted with caution. In particular, telemetry devices were affixed to polar bears each spring between mid-March and late-May opportunistically, with polar bears close to logistical bases located along the Arctic coast more likely to be tagged than polar bears far from the coastline \citep{Wilson2014, Wilson2016}. Additionally, the majority of units were affixed to females, as male polar bears require specialized radio transmitters. Thus, it is possible that the way in which data were gathered may be confounding our inference about the start of the ``summer'' season.

The posterior distribution for $\tau$ had a median value of approximately 90km, suggesting that polar bears prefer habitat within about 180km of the sea ice edge or coastline during the ``summer'' season on average. In comparison, \cite{Durner2004} concluded that polar bears selected habitat near the sea ice edge during July 24--May 30, with the strongest effect occurring during July 24--November 15 when polar bears showed a preference for habitat within about 100km. \cite{Durner2004} also showed that, from November 16--May 30 (approximately the period of maximum sea ice extent), polar bears continued to show a weakened preference for habitat near the sea ice interface, but at a larger range of distances. The relevant findings in \cite{Durner2009} were similar; polar bears showed a preference for habitat within approximately 200km of the sea ice edge. \cite{Durner2009} also found that, during the period of maximum sea ice extent, polar bears showed a preference for habitat within approximately 200km of coastlines. Thus, our conclusions about the timing and relevant spatial range of the effect of sea ice edge and coastline are consistent with those discussed in previous studies. 

\section{Discussion}
Our novel approximation method based on the linearization of a potentially time-varying spatial feature, $\mcM(t)$, reduces the computational burden of fitting models in the common RSF framework, allowing researchers increased flexibility in the types of RSFs they can specify in mechanistic models for movement. We demonstrated our approach in an application involving the movement of polar bears as they responded to seasonal shifts in sea ice during 2012--2016. Recently, there has been a significant increase in research focused on the so-called ``greenwave'' hypothesis \citep[e.g.,][]{Aikens2017}, which posits that herbivorous animals align their movement with bands of high-quality forage that shift throughout spring as different elevations and latitudes experience phenological changes. Our modeling approach represents a way to validate the hypothesis if information about the shape of the greenwave is known, or potentially estimate the location of the posited band of high quality forage based on the observed movement patterns of herbivores. 

Several models for animal movement that seek to elucidate the complex relationship between individuals and their habitat have been proposed in the literature. Many recent models for movement are not explicitly framed in terms of RSFs, yet share important connections with our proposed methodology. A few important examples are \cite{Hanks2011, Hanks2015}, and \cite{Buderman2018} who used the same multiple-imputation based two-stage approach to model individual responses to landscape features in a discrete-space framework. While \cite{Hanks2011, Hanks2015} required a third stage of analysis to obtain population-level inference, \cite{Buderman2018} used a hierarchical model construction similar to our approach that permits joint inference about individual and population-level parameters of interest, albeit at the cost of additional computation time. Additionally, \cite{Johnson2008} and \cite{Scharf2017} modeled individual-level movement using a continuous-space framework that is able to capture a tendency of individuals to ``drift'' toward particular landscape features. Importantly, while \cite{Hanks2011, Hanks2015}, \cite{Scharf2017}, and \cite{Buderman2018} allowed for time-varying responses to landscape features (analogous to the way we modeled a seasonal response of polar bears to the habitat described by $\mcM(t)$), none of these approaches explicitly accounted for dynamic landscape features that change in shape and location through time.

The linearization approximation methodology can also be extended to higher-dimensional spaces and features. For instance, in marine environments, RSFs based on two-dimensional features, such as isotherms, may be locally approximated using rank-deficient Gaussian distributions corresponding to infinite planes. One-dimensional features in three-dimensional spaces, such as wind or ocean currents, can also be approximated with improper distributions (see Supporting Information~E for mathematical details). In practice, researchers will need to evaluate the appropriateness of a linearization approximation; however, in many cases, our methodology offers a way to include complex drivers of movement that might otherwise have been computationally inaccessible.

%  The \backmatter command formats the subsequent headings so that they
%  are in the journal style.  Please keep this command in your document
%  in this position, right after the final section of the main part of
%  the paper and right before the Acknowledgements, Supplementary Materials,
%  and References sections.

\backmatter

%  This section is optional.  Here is where you will want to cite
%  grants, people who helped with the paper, etc.  But keep it short!

\section{Supplementary Materials}
Web Appendices A--E, which provide details related to the definition of the dynamic landscape feature (A), MCMC implementation (B), sub-population spatial boundary (C), simulation study (D), and extensions of the linearization approximation to higher dimensions (E) are available with this paper at the Biometrics website on Wiley Online Library.

\section*{Acknowledgments}
The authors thank Franny Buderman, Perry Williams, and Christopher Peck for insight that improved the article. Any use of trade, firm, or product names is for descriptive purposes only and does not imply endorsement by the U.S. Government.

The views of the USFWS authors in this publication are solely those of the USFWS authors, and do not necessarily represent the views of the USFWS. This article has been peer-reviewed and approved by USGS under their Fundamental Science Practices policy (http://pubs.usgs.gov/circ/1367). This research was permitted under the Marine Mammal Protection Act and Endangered Species Act under U.S. Fish and Wildlife Service permits (MA046081, MA 690038) and followed protocols approved by Animal Care and Use Committees of the U.S. Fish and Wildlife Service and U.S. Geological Survey.

The authors acknowledge support for this research from  NSF DMS 1614392 and USFWS G17AC00068.

%  If your paper refers to supplementary web material, then you MUST
%  include this section!!  See Instructions for Authors at the journal
%  website http://www.biometrics.tibs.org

% Supporting Information~A, referenced in Section~\ref{sec:model}, is available with
% this paper at the Biometrics website on Wiley Online
% Library.\vspace*{-8pt}

%  Here, we create the bibliographic entries manually, following the
%  journal style.  If you use this method or use natbib, PLEASE PAY
%  CAREFUL ATTENTION TO THE BIBLIOGRAPHIC STYLE IN A RECENT ISSUE OF
%  THE JOURNAL AND FOLLOW IT!  Failure to follow stylistic conventions
%  just lengthens the time spend copyediting your paper and hence its
%  position in the publication queue should it be accepted.

%  We greatly prefer that you incorporate the references for your
%  article into the body of the article as we have done here
%  (you can use natbib or not as you choose) than use BiBTeX,
%  so that your article is self-contained in one file.
%  If you do use BiBTeX, please use the .bst file that comes with
%  the distribution.  In this case, replace the thebibliography
%  environment below by
%

\bibliographystyle{temp}
\bibliography{../../polar_bears.bib}

% \begin{thebibliography}{}

% \bibitem{ } Cox, D. R. (1972). Regression models and life tables (with
% discussion).  \textit{Journal of the Royal Statistical Society, Series B}
% \textbf{34,} 187--200.

% \bibitem{ }  Hastie, T., Tibshirani, R., and Friedman, J. (2001). \textit{The
% Elements of Statistical Learning: Data Mining, Inference, and Prediction}.
% New York: Springer.

% \end{thebibliography}

%  To get the journal style of heading for an appendix, mimic the following.

\label{lastpage}

\end{document}